\documentclass[aps]{revtex4}
\newcommand{\bb}{\begin{eqnarray}}
\newcommand{\ee}{\end{eqnarray}}
\begin{document}
\title{Complex fermion mass term, regularization and CP violation}
\author{P. Mitra}\email{parthasarathi.mitra@saha.ac.in}
\affiliation{Saha Institute of Nuclear Physics,\\ 1/AF Bidhannagar,\\
Calcutta 700064, India}

\begin{abstract}
{It is well known that the CP violating 
theta term of QCD can be converted to a
phase in the quark mass term. However, a theory with
a complex mass term for quarks can be regularized so as not to
violate CP, for example through a zeta function. 
The contradiction is resolved through
the recognition of a dependence on the regularization or measure.
The appropriate choice of regularization is discussed
and implications for the strong CP problem are pointed out.}
\end{abstract}
\maketitle

\section*{Introduction}

In the standard model, the symmetry breaking in the electroweak sector
leaves a $\gamma^5$ phase in the mass term it produces for quarks,
and this phase is usually considered to
break parity and time-reversal in the strong
interaction sector. Indeed, it was shown many years ago \cite{baluni}
that the colour gauge field term involving the vacuum angle $\theta$ can be converted to a 
$\gamma^5$ phase in the quark mass term. Thus for a complete description
of the expected  parity and time-reversal violation, 
it is common to introduce an effective
$\bar\theta$, which contains $\theta$ together with a contribution
from the phase in the quark mass term. 
This expected violation has been a puzzle -- called the strong CP problem -- 
because it is {\it not} experimentally observed \cite{kim}. 
There have been attempts to explain why it should not be seen, {\it e.g.,}
by \cite{sachs}, but the situation is unclear.
For a long time the problem has been sought to be solved by schemes
involving the existence of particles called axions, but these have not been
detected \cite{ax} in spite of extensive searches. This necessitates
further investigation of the problem.

We start by noting that even $\bar\theta$ is not the whole story:
to know what happens in the quantized theory, it
is necessary to specify a regularization. It is clear that in general
the amount of CP violation depends on the regularization, so that there is a
scope for the use of a third term signifying the contribution of
the regularization to the actual $\bar\theta$. Of course, there may be reasons 
for choosing some regularizations over some others, and this may take away some
of the freedom in the third term.

If an explicit regularization like Pauli-Villars is used, the regulator fields
can have their own chiral phase, and this is known to contribute
to $\bar\theta$. It was demonstrated in \cite{bcm} that this phase can be chosen
to conserve CP in the theory of a fermion having a complex mass term. 
At first sight this might appear to contradict \cite{baluni} where
the complex mass term was shown to be obtainable from a CP violating term.
Indeed, if a Pauli-Villars regularization is
used with zero phase in the mass terms of the regulator fields,
the CP violation expected according to \cite{baluni} is recovered \cite{dine}.
The reconciliation comes from the observation that different regularizations
are implicit in the two cases and the total CP violation depends on the
regularization. 

The above observation may lead one to ask whether both the phases used
for the regulator fields are acceptable {\it i.e.} whether the CP violation
expected according to \cite{baluni} can be legitimately removed. In an
attempt to clarify the issue, we
shall first explore the theory involving a complex fermion mass term
in a more direct regularization, namely one using the zeta function.
It turns out, without any tampering of phases,
that there is no resultant CP violation.

Thereafter, we shall study a class of choices of the measure of fermion 
functional integration, where the measure will be treated in a formal
way and implicitly involve a regularization. The total CP violation will
depend on the choice. A symmetry principle places some restriction
on the measure: the CP violating effects of the mass term and the measure
cancel out in this situation. This has a bearing on the strong CP problem
\cite{kim}: it essentially ceases to exist. 

\section*{Zeta function regularization}
The standard fermion mass term (with zero phase) is 
\bb
-\int\bar\psi m\psi.
\ee
But there is a nonzero phase:
\bb
S_M=-\int\bar\psi m\exp (i\alpha\gamma^5)\psi.\label{alpha}
\ee
The Dirac operator 
\bb
i\not\partial+\not{A}-m\exp (i\alpha\gamma^5)
\ee
is not hermitian. The positive operator needed for the zeta
function (see {\it e.g.,} \cite{zeta}) is constructed as
\bb
\Delta=[i\not\partial+\not{A}-m\exp (i\alpha\gamma^5)]^\dagger
[i\not\partial+\not{A}-m\exp (i\alpha\gamma^5)].
\ee
It is straightforward to see that for antihermitian $\gamma$-matrices,
as appropriate for euclidean spacetime,
\bb
\Delta=-(i\not\partial+\not{A})^2+m^2,
\ee
which is {\it independent} of the phase $\alpha$.
The zeta function that appears involves a parameter $s$,
\bb
\zeta(s,\Delta)\equiv{\rm Tr}(\Delta^{-s}),
\ee 
and the fermion determinant is defined in the limit of $s\to0$ as
\bb
-\frac12\zeta'(0,\Delta)-\frac12\ln\mu^2\zeta(0,\Delta).
\ee
This is independent of $\alpha$ and depends on the gauge fields
through the operator $\Delta$. It is the same determinant as in the
case where the phase in the mass term vanishes and is therefore
invariant under CP transformations of the gauge field $A$.
This confirms that a phase in the quark mass term does not cause any
CP violation in quantum chromodynamics if a suitable regularization
is used \cite{bcm}.

\section*{A class of formal fermion measures}

Let us consider a 
fermion action with a real mass term and invariant under parity $P$:
\bb
S[\psi,\bar\psi, A]=S[P\psi,\bar\psi P, A^P].
\ee
Here $A$ is the gauge field and  $A^P$ its parity-transformed form.
If we apply a chiral transformation $\exp (i\alpha\gamma^5/2)$ on the fields,
the mass term (\ref{alpha}) can be generated.
More generally, we apply a chiral transformation $\chi=\exp (i\beta\gamma^5/2)$.
Then the action will change, and will no longer be invariant under $P$,
but it will clearly be invariant under a new transformation:
\bb
S_\chi[\psi,\bar\psi, A]&\equiv& S[\chi\psi,\bar\psi\chi, A]\nonumber\\
&=&S[P\chi\psi,\bar\psi\chi P, A^P]\nonumber\\
&=&S[\chi\chi^{-1}P\chi\psi,\bar\psi\chi P\chi^{-1}\chi, A^P]\nonumber\\
&=&S_\chi[\chi^{-1}P\chi\psi,\bar\psi\chi P\chi^{-1}, A^P].\label{P}
\ee
This symmetry simply involves $P$ rotated by $\chi$ \cite{bcm}. 

What about the fermion regularization or measure? For the action $S$,
the conventional fermion measure, with an implicit
regularization, is invariant under $P$:
\bb
d\mu[\psi,\bar\psi, A]=d\mu[P\psi,\bar\psi P, A^P].
\ee
But this $d\mu$ will not in general respect the symmetry (\ref{P}) as
gauge fields are present and there is a chiral anomaly: 
\bb
d\mu[\psi,\bar\psi, A]\neq
d\mu[\chi^{-1}P\chi\psi,\bar\psi\chi P\chi^{-1}, A^P].
\ee
However, we can manufacture a new measure
\bb
d\mu_\chi[\psi,\bar\psi, A]&\equiv& d\mu[\chi\psi,\bar\psi \chi, A]\nonumber\\
&=&d\mu[P\chi\psi,\bar\psi \chi P, A^P]\nonumber\\
&=&d\mu_\chi[\chi^{-1}P\chi\psi,\bar\psi\chi P\chi^{-1}, A^P],
\ee
which formally has the same symmetry (\ref{P}) as $S_\chi$.

Thus there is already a choice of two measures $d\mu$ and $d\mu_\chi$
available for use with the action $S_\chi$. From the above construction
it should be clear that the measure has a parameter $\beta$ and this
parameter may be chosen to be different from the $\alpha$ in the action,
so that for each phase in the mass term, there is a scope for
a one-parameter class of variations of the measure.

These constructions work in the presence of gauge fields,
and it should be noted that anomalies have been implicitly
taken into account here: it is only because
of anomalies that the measure $d\mu_\chi$ depends on $\chi$.

The total CP violation gets contributions from the
vacuum angle $\theta$ (if any), the phase $\alpha$ 
in the fermion mass term and the
phase $\beta$ in the definition of the measure $d\mu_\chi$:
\bb
\bar\theta=\theta-\alpha+\beta.
\ee
The first two terms on the right hand side have been
familiar for a long time, and it has also been
known that the phase of a regulator term like the
Pauli-Villars mass term should contribute to the CP
violation, but this form of the equation with three
terms on the right hand side is not so well known.

It is interesting to note that one may take $\alpha$
and $\beta$ to be equal:
there exists a choice of the measure that
respects the symmetry (\ref{P}) of the complex fermion action. 
CP is then broken only by the vacuum angle $\theta$ 
in the gauge field action. 
However, at this stage, other choices of measure are also legitimate.

\section*{Choice of measure}
Is there a preferred measure? For a theory with a real mass
term, it would be instinctive to take {\it the} parity
invariant measure. This is what we have denoted by $d\mu$.
When the mass term is complex, one could of course continue
to use the same measure, but as we have observed, there are
other measures which can be obtained by chiral rotation.
When the action is $S_\chi$, it has a specific parity
symmetry and we have seen that a specific measure $d\mu_\chi$
shares the same symmetry. It is customary to demand that
when an action is regularized, the regularization should
have the symmetry of the action and this means that the measure
$d\mu_\chi$ should be preferred when the fermion action is
$S_\chi$, which simply means that one should impose
\bb
\beta=\alpha.
\ee
If this is done, the fermion sector as a whole does not violate CP.
The only CP violation comes from the gauge sector:
\bb
\bar\theta=\theta.
\ee
This is exactly as in the case of the direct zeta function
regularization of the fermion determinant, which does not
however quite belong to this measure formalism.
\section*{Conclusion}
 
We have first shown that in the zeta function regularization
there is no CP violation from the phase of a complex fermion mass term.
Then we have investigated the possibility of using chirally
rotated regularizations/measures containing chiral phases $\beta$.
Such phases in the fermion
measure have been overlooked in the literature, {\it i.e.,}
the real measure has been used
even when the fermion mass term is complex. 

There is no direct conflict in conventional
perturbation theory between the measures $d\mu_\chi$ with different values of $\chi$. 
As is well known, these measures
can be converted to one another, with an alteration only in the
$F\tilde F$ term of the action. But this term is a total derivative, 
and its integral is not visible
in conventional perturbation theory.  

There is also no contradiction of such phases in the measure with gauge
invariance or with our understanding of the anomaly. If a chiral transformation is
carried out, the phase of the mass term changes, and {\it if
the measure is held fixed}, the standard anomaly emerges unscathed from the
Jacobian because chiral transformations commute with one another. 
Of course, if the measure is altered, this 
produces another Jacobian which may cancel the earlier Jacobian.
In our notation, if $\alpha$ is altered, $\bar\theta$ changes,
but if $\beta$ is also altered, the change in $\bar\theta$ may be cancelled.

One may ask why the measure should be changed.
Changing the {\it mass} does not call for a change in the functional
measure of a field, but changing its {\it phase} does result in a change of the  
symmetry transformations of the fermion action --  not of the
symmetry group -- and this is what suggests a change in the measure. 
The imposition of the symmetry of the action on the regularization or measure 
is a very widely used principle which can be followed to fix the 
measure. Without this principle, any one of our continuous family of
measures with different values of the phase can be used, and the resulting
CP violation depends on this arbitrary phase as well as the phase in the
mass term. But if the classical symmetry of the fermion action is
taken seriously, it forces the specific choice
$\beta=\alpha$ for the phase in the measure. 

This choice has a serious consequence.
The chiral phase in the mass term is balanced by the chiral phase
in the measure in such a way that their CP violating effects cancel out.
Thus the chiral phase in the mass term
has no net physical effect, as in the case of the zeta function regularization.
One can say that the usual $\bar\theta$ gets a contribution
from the fermion sector only when there is a mismatch between the parity
symmetry of the action and the parity symmetry
of the measure. If the measure has the same 
parity symmetry as the action there is no breaking of
CP from the fermion sector.
A CP violation could still come from the vacuum angle $\theta$,
but it may be set equal to zero to satisfy experimental observations:
there is no unnatural fine-tuning involved in this. 
Thus there is really no strong CP problem \cite{kim} and no need for 
axions which have not been detected anyway \cite{ax}.


\end{document}